\documentclass{jpsj-suppl}
\usepackage{txfonts} 
\usepackage{bbm}
\usepackage{bm}
\usepackage{epstopdf}
\usepackage{color}

\title{The nature of near-threshold $XYZ$ states}

\author{Martin \textsc{Cleven}$^{1}$, Feng-Kun \textsc{Guo}$^{2,3}$, Christoph
\textsc{Hanhart}$^{4}$, Qian \textsc{Wang}$^{4}$ and Qiang \textsc{Zhao} $^{1}$}

\inst{$^{1}$Institute of High Energy Physics and Theoretical
       Physics Center for Science Facilities, Chinese Academy of Sciences,
       Beijing 100049, China \\
$^{2}$Helmholtz-Institut f\"ur Strahlen- und Kernphysik
       and Bethe Center for Theoretical Physics, Universit\"{a}t Bonn, D-53115
       Bonn, Germany\\
 $^{3}$State Key Laboratory of Theoretical Physics,
Institute of Theoretical Physics, CAS,\\ Beijing 100190, China \\
 $^4$ Institut f\"{u}r Kernphysik and  Institute for Advanced Simulation,
       Forschungszentrum J\"{u}lich, D--52425 J\"{u}lich, Germany}

\email{q.wang@fz-juelich.de}

\recdate{October 1, 2015}

\abst{ We demonstrate that the recently observed $X$, $Y$, $Z$ states cannot be
purely from kinematic effect. Especially the narrow near-threshold structures in
elastic channels call for nearby poles of the $S$-matrix which are qualified as
states.
We propose a way to distinguish cusp effects from genuine states and demonstrate
that (not all of) the recently observed $X$, $Y$, $Z$ states cannot be purely from
kinematic effects.
 Especially, we show that the narrow near-threshold structures in elastic
 channels call for nearby poles of the $S$-matrix, since the normal kinematic
 cusp effect cannot produce that narrow structures in the elastic channels in
 contrast to genuine $S$-matrix poles.
 In addition, it is also discussed how spectra can be used to distinguish
 different scenarios proposed for the structure of  those poles, such as
 hadro-quarkonia, tetraquarks and hadronic molecules. The basic tool employed is
 heavy quark spin symmetry. }

\kword{Exotic mesons, Meson-meson interactions}

\begin{document}
\maketitle

\section{Introduction}

In the last decade  many narrow structures were observed which do not fit into
the well-established quark model in both charmonium and bottomonium sectors.
Most of them are located close to nearby $S$-wave open flavor thresholds. For
instance, in the charmonium sector  $X(3872)$ and $Z_c(3900)$ are close to the
$DD^*$~(here and in the following, one of the two open-flavor mesons contains a
heavy quark and the other contains a heavy anti-quark) threshold and $Z_c(4020)$
 is close to the $D^*D^*$, while in the bottomonium sector the $Z_b(10610)$ and
$Z_b(10650)$ are close to $BB^*$ and $B^*B^*$, respectively. Due to their
proximity to the thresholds,  various groups conclude that those structures are
simply kinematical effects~\cite{Swanson:2014tra,Swanson:2015bsa} (and
references therein) which occur near every $S$-wave threshold.
In contrast to this there are many publications where the experimental
signatures are interpreted as states with various suggestions for the underlying
structure. The most prominent proposals are
hadro-quarkonia\cite{Voloshin:2007dx,Dubynskiy:2008mq}, hybrids
\cite{Zhu:2005hp,Kou:2005gt,Close:2005iz},
tetraquarks\cite{Maiani:2014aja,Faccini:2013lda,Hogaasen:2005jv,Buccella:2006fn,Guo:2011gu,Hogaasen:2013nca,Stancu:2009ka} and hadronic
molecules~\cite{Tornqvist:2004qy,Fleming:2007rp,Thomas:2008ja,Ding:2008gr,Lee:2009hy,Dong:2009yp,Stapleton:2009ey,
Gamermann:2009fv,Mehen:2011ds,Bondar:2011ev, Nieves:2011vw,Nieves:2012tt,
Wang:2013cya,Wang:2013kra,Guo:2013sya,Guo:2013zbw,
Mehen:2013mva,He:2013nwa,Liu:2014eka}.

This contribution consists of two parts. In the first part we  demonstrate that
although there is always a cusp at the opening of an $S$-wave threshold, it
cannot produce a narrow pronounced structure in the elastic channel  (the
channel where the final state agrees to the channel that produces the cusp) from
a perturbative rescattering~\cite{Schmid:1967,Guo:2014iya}.
As a consequence of this, the narrow structures in elastic channels as observed
in experiments necessarily call either for nonperturbative interactions between
heavy mesons that lead to poles of the $S$-matrix or call for poles formed on
the quark level.
Having established that we need to talk about genuine states, in the second part
of this contribution, we  employ heavy quark spin symmetry (HQSS)  to
distinguish three different dynamical models, i.e. hadro-quarkonia, tetraquarks
and hadronic molecules based solely on their mass spectra~\cite{Cleven:2015era}.

\section{Kinematic effect or $S$-matrix pole?}
\begin{figure}[t] \vspace{0.cm}
\begin{center}
\hspace{2cm}
\includegraphics[width=0.9\linewidth]{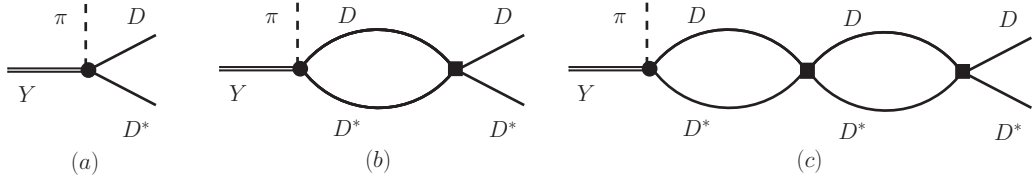}
\caption{The tree-level (a), one-loop (b) and two-loop (c) Feynman diagrams for $Y(4260)\to \pi D D^*$.}
 \label{fig:Fey1}
\end{center}
\end{figure}
In this section, we analyse the existing data of $Z_c(3900)$ observed in
$Y(4260)\to \pi D D^*$ to illustrate our argument about how to distinguish
the presence of
an $S$-matrix pole from a purely kinematic effect. However, it should be clear
that the conclusion is more general and can be applied to all  narrow structures
near  $S$-wave thresholds such as those $XYZ$ states mentioned above.  To
illustrate this point, we do not aim at  field theoretical rigor  but use a
separable interaction for all vertices,
\begin{eqnarray}
{\cal L}_I&=&g_Y  \pi (D\bar D^*_\mu)^\dagger Y^\mu + \frac{C}{2} (D\bar D^*)^\dagger(D\bar D^*)+\cdots,
\end{eqnarray}
and a Gaussian regulator
\begin{equation}
f_\Lambda(\vec p\, ^2) = \exp\left(-2\vec p\, ^2/\Lambda^2\right) \ ,
\label{eq:ff}
\end{equation}
for each loop.  Here $Y^\mu$, $D$, $D^*_\mu$ and $\pi$ are the fields for the $Y(4260)$, $D$, $D^*$ and $\pi$, respectively. The coupling $g_Y$ is the source term strength and $C$ is the $DD^*$ rescattering strength. It is the size of this parameter that will eventually allow
us to decide whether the rescattering effect is perturbative or not. In the
regulator, $\vec{p}$ is the three momentum of the $D$-meson in the
center-of-mass (c.m.) frame of the $DD^*$ system and $\Lambda$ is the cut-off
parameter. Therefore, the loop function reads
 \begin{eqnarray}
G_\Lambda(E)&=&\int \frac{d^3 q}{(2\pi)^3}\frac{f_\Lambda(\vec q\, ^2)}{E-m_1
-m_2 -\vec q\, ^2/(2\mu)} \ =\frac{ \mu\Lambda }{(2\pi)^{3/2}} + \frac{\mu k}{2\pi}
e^{-2k^2/\Lambda^2} \left[ \text{erfi}\left(\frac{\sqrt{2}k}{\Lambda}\right) - i
\right]
\end{eqnarray}
where $k = \sqrt{2\mu (E-m_1-m_2)}$, and
$
\text{erfi}(z) = 2/({\sqrt{\pi}})
\int_0^z e^{t^2}dt
$.

\begin{figure}[t]
\centering
\includegraphics[width=1.0\linewidth]{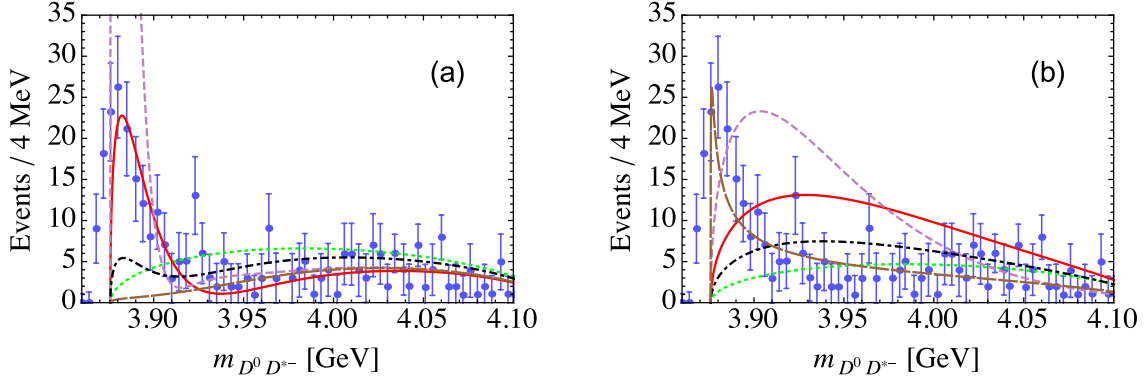}
\caption{The $DD^*$ invariant mass distribution in $Y(4260)\to \pi DD^*$. The data are from Ref.\cite{Ablikim:2015swa} and the green dotted, red solid, pink dashed and brown long dashed curves are the contributions from the tree level, full one-loop, full two-loop and the sum of all the loops, respectively.  The dot-dashed line shows the full one-loop result with the strength of the rescattering requested to be small enough to justify a perturbative treatment as described in the text. Figure (a) and Figure (b) show the results with the full one-loop and the sum of all the loops fitted to the experimental data, respectively. The events below the $DD^*$ threshold are from the energy resolution that is not included in the theoretical
calculations.}
\label{fig:DDstar}
\end{figure}

For the kinematic explanations of these $XYZ$ states \cite{Swanson:2014tra,Swanson:2015bsa}, it was claimed that the sum of tree-level (Fig.\ref{fig:Fey1} (a)) and one-loop (Fig.\ref{fig:Fey1} (b)) diagram can describe the invariant mass distribution quantitatively.
In other words, the claim is that the rescattering process is perturbative and one does not need to add also the higher order  loops.
We  confirm the result of the
calculation by employing the tree-level diagram plus the one-loop diagram  to fit the new data from BESIII~\cite{Ablikim:2015swa}  as shown by the red solid curve in Fig.~\ref{fig:DDstar} (a). The fitted parameters are
\begin{eqnarray}
g_Y=40.01~\mathrm{GeV}^{-3/2},\quad C=187.37~\mathrm{GeV}^{-2},\quad \Lambda=0.257~\mathrm{GeV}.
\end{eqnarray}
The results shown here are updated compared to our previous results \cite{Guo:2014iya}
where the used data did not unambigously identify the $D D^*$ final state.
In contrast to this for the new data the final state was measured exclusively.
This leads to a moderate decrease of the
data at the higher energies.
We find that this decrease requires an increase in the rescattering strength, without changing the general reasoning. In particular,
also from this analysis it follows that one has to sum all the bubble loops up
to infinite order for a consistent treatment. To illustrate this point we plot
the full two-loop contribution (Fig.~\ref{fig:Fey1} (a) + (b) + (c)) as  dashed
curve in Fig.~\ref{fig:DDstar} (a):
 The deviation of this curve from the full one-loop contribution is much
larger than what is allowed in a perturbative scheme.
In addition: a summation of the bubble leads to a bound state with a 9~MeV
binding energy. This clearly illustrates that restricting oneself to a
single rescattering is not self-consistent.

Turning the argument around, we can see that the perturbative requirement,
namely that the two-loop contribution is at most half of the one-loop
contribution, cannot produce the pronounced near-threshold structure as shown by
the dot-dashed line in Fig.~\ref{fig:DDstar} (a).  The crucial point of this
analysis is that in the elastic channel the contribution from the source
term (Fig.~\ref{fig:Fey1} (a)),  as shown by the dotted curve in
Fig.~\ref{fig:DDstar} (a), can be disentangled from the $DD^*$
rescattering process Fig.~\ref{fig:Fey1} (b). The former one is fixed by the $DD^*$ invariant mass
distribution above $\sim 3.94~\mathrm{GeV}$ and the latter one is used to
explain the near-threshold structure.

As discussed above, since the narrow near-threshold structure means a
nonperturbative rescattering process, one needs to sum all the bubble loops to
infinite order. Doing this to fit the data we find for the parameters
\begin{eqnarray}
g_Y=33.85~\mathrm{GeV}^{-3/2},\quad C=20.09~\mathrm{GeV}^{-2},\quad \Lambda=0.89~\mathrm{GeV}.
\end{eqnarray}
The fit results are shown in  Fig.~\ref{fig:DDstar} (b).
With the above fitted parameters, we find a pole below the $DD^*$ threshold
with the binding energy $0.59~\mathrm{MeV}$. This demonstrates that a narrow
pronounced near-threshold structure in the elastic channel requires a pole in
the $S$-matrix and cannot be produced from kinematic effects alone.

Our reasoning is in contrast to that of Ref.~\cite{Swanson:2015bsa}, which is to
our knowledge the only work so far that claims a kinematic origin for the
near--threshold structures and also looks at the elastic channel. It should be
stressed that in this work, while the structures in the inelastic channels are
explained as cusps, those in the elastic channels come solely from the form
factors employed. In this way our argument presented above is evaded. However,
from our point of view the problem of this explanation is that it appears
unnatural to explain both the narrow structure as well as the higher energy tail
in the elastic channels as form factors because this calls for the simultaneous
presence of drastically different length scales in the production vertex. We
therefore do not regard the mechanism of Ref.~\cite{Swanson:2015bsa} as a
plausible explanation for the near threshold structures observed. Thus, for the
rest of these proceedings we regard it as established that the $XYZ$ structures
are states and discuss a certain proposal, how their internal structure could be
disentangled experimentally.

\section{Different scenarios for the $XYZ$ states}\label{sec:XYZ}
Having established that the $XYZ$ states  require a near-threshold pole
structure in the elastic channel we now discuss how to distinguish different
models for these poles such as hadro-quarkonia, tetraquarks and hadronic
molecules~\cite{Cleven:2015era}.

\subsection{Hadro-quarkonia}
The hadro-quarkonium picture was proposed by
Voloshin~\cite{Voloshin:2007dx,Dubynskiy:2008mq} based on the fact that several
exotic candidates mainly decay into a heavy quarkonium plus  light hadrons.
Examples are  the $Y(4260)$ discovered in $J/\psi\pi\pi$ channel, the
$Z_c(4430)$  discovered in $\psi^\prime\pi$ channel, and the $Y(4360)$ and
$Y(4660)$  observed in $\psi^\prime\pi\pi$ channel. The basic idea is that the
$XYZ$ states contain a compact heavy-quarkonium core surrounded by a
light-meson cloud. Based on the recent measurement that the
cross sections for $J/\psi\pi^+\pi^-$ and $h_c\pi^+\pi^-$ at
$4.26~\mathrm{GeV}$ and $4.36~\mathrm{GeV}$ in $e^+e^-$ collisions are of
similar size, Li and Voloshin include HQSS breaking to describe the $Y(4260)$ and $Y(4360)$ as a mixture of two hadro-charmonia~\cite{Li:2013ssa}:
\begin{eqnarray}
  Y(4260) = \cos\theta\,\psi_3 - \sin\theta \, \psi_1\, ,
 \qquad  Y(4360)  = \sin\theta \,\psi_3 + \cos\theta
 \,\psi_1\, ,
 \label{eq:Ymixing}
\end{eqnarray}
where $\psi_1\sim (1^{+-})_{c\bar c}\otimes
(0^{-+})_{q\bar q}$ and  $\psi_3\sim (1^{--})_{c\bar c}\otimes (0^{++})_{q\bar
q}$ are the wave functions of the $J^{PC}=1^{--}$ hadro-charmonia with a
$1^{+-}$ and $1^{--}$ $c\bar c$ core charmonium, respectively. Since the leading order interaction between the heavy core and the light meson cloud is not dependent on the spin of the heavy core, spin partners for the heavy  hadro-quarkonia can be identified by
replacing the core with the corresponding spin partner~\cite{Cleven:2015era}.
\begin{figure}[t]
\centering
\includegraphics[width=0.5\linewidth]{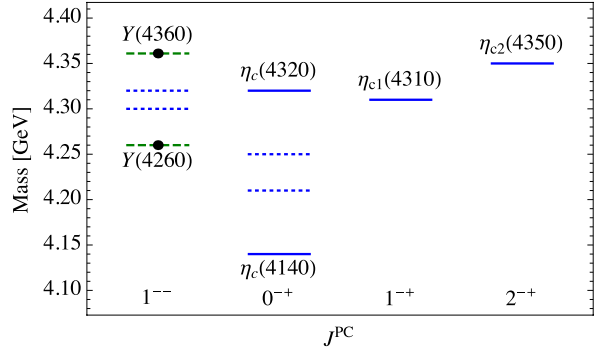}
\caption{(color online). The full spectroscopy of hadro-charmonia based on the assumption that $Y(4260)$ and $Y(4360)$ are the mixed states of two $1^{--}$ hadro-charmonia\cite{Li:2013ssa}. The dotted lines are the unmixed states. The masses of $Y(4260)$ and $Y(4360)$ are the inputs as shown by the dashed green lines.}
\label{fig:hadrocharmonium}
\end{figure}

Accordingly, the spin partners of $Y(4260)$ and $Y(4360)$ are found by replacing the $\psi'$ in the core of $\psi_3$ by $\eta_c'$
and replacing the $h_c$ in the $\psi_1$ state by any of the three $\chi_{cJ}$ states.
The relative mixing amplitude can be obtained by constructing an $CP$-odd operator with the HQSS breaking~\cite{Cleven:2015era}
\begin{equation}
  \mathcal{O}_\text{mixing} = \frac14\langle
  \vec{\chi}^{\,\dag}\cdot\vec{\sigma}J' \rangle + \text{h.c.}
  = \vec{h}_c^\dag \cdot \vec{\psi}' + \sqrt{3}\,\chi_{c0}^\dag\eta_c' +
  \text{h.c.}.
\end{equation}
Hence, the mixing amplitude of the pseudoscalar sector is larger by a factor $\sqrt{3}$ than that in the vector sector.
The resulting masses of these hadro-charmonia are shown in Fig.~\ref{fig:hadrocharmonium} as an illustration.
We notice that the interpretation of $Y(4260)$ and $Y(4360)$
as mixed hadron-charmonia implies the existence of two states with $J^{PC}=0^{-+}$.

The search for these partners will provide more insights into the nature of the
two $Y$ states.
For the two pseudoscalar states, i.e. $\eta_c(4140)$ and $\eta_c(4320)$, the
large mixing amplitude leads to masses that allow  both of them to decay into
 $\eta_c^{(\prime)}\pi\pi$ and $\chi_{c0}\pi\pi$. Because of their quantum
 numbers the states cannot be produced directly in $e^+e^-$ collisions. An
 alternative way for searching for them is in $B$ meson decays, e.g. $B^\pm\to
 K^\pm \eta_c^{(\prime)}\pi^+\pi^-$ as suggested in
Ref.~\cite{Guo:2009id} for the search of the spin partner of the $Y(4660)$.
Another way to search for these two pseudoscalars is the radiative decays of
$Y(4260)$ and $Y(4360)$ via their $\psi'$ component. Since the branching ratio
of $\psi'\to\gamma\chi_{c0}$ is two orders of magnitude larger than that for
$\psi'\to\gamma\eta_c'$, one could expect to observe these two states in  the
$e^+e^-\to\gamma \chi_{c0}\eta$ process at  center-of-mass energies around the
masses of the $Y(4260)$ and the $Y(4360)$. The states $\eta_{c1}(4310)$ and
$\eta_{c2}(4350)$ can be searched for in analogous processes in the decays of
$Y(4360)$ with the $\chi_{c0}$ in the final state replaced by the $\chi_{c1}$
and $\chi_{c2}$, respectively. The search for the above states can be performed at BESIII 
or a future high-luminosity super tau-charm factory.

\subsection{Tetraquarks}
Among the different tetraquark scenarios  we here focus on  the compact
diquark-antidiquark states suggested by Maiani et al.~\cite{Maiani:2014aja} for
simplicity. It is a straightforward extension of
the quark model.
Since the four-quark system is bound by the effective gluon exchanges, the
isoscalar tetraquark is mostly degenerate with the isovector tetraquark  similar
to the $\rho$--$\omega$ degeneracy in the light meson sector.

In this model, the mass of a tetraquark is given by~\cite{Maiani:2014aja,Cleven:2015era}
\begin{equation}
\label{eq:2}
M=M_{00}+B_c\frac{L(L+1)}{2}+a[L(L+1)+S(S+1)-J(J+1)]+
\kappa_{cq} \left[s(s+1)+\bar{s}(\bar{s}+1)-3\right] \ ,
\end{equation}
with $s$ and $\bar{s}$ the total spin of diquark and antidiquark system, $S$ the total spin, $L$ the relative orbital angular momentum between diquark and antidiquark system, $J$ the total angular momentum.
The parameters $B_c$, $a$ and $\kappa_{cq}$ are defined such that they are
positive to fit to the masses of selected states.
As a result, the mass of the tetraquarks increases with increasing $L$ and $S$, but decreases for growing $J$, which is a rather unusual feature for composite systems.
\begin{figure}[t]
\centering
\includegraphics[width=0.9\linewidth]{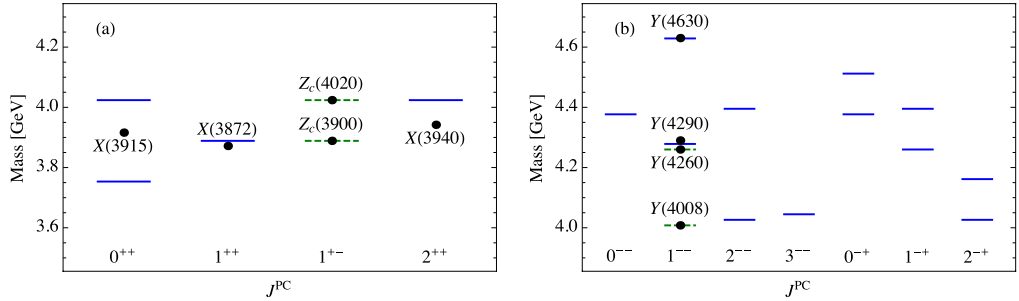}
\caption{ Panel (a) and (b) stand for the $S$-wave and $P$-wave tetraquark spectroscopy in the charmonium sector, respectively. The input and prediction are denoted by the green dashed and blue solid lines, respectively.}
\label{fig:tetraquarks}
\end{figure}

As discussed above, the most striking feature  is a very rich spectroscopy emerging in the tetraquark model, as shown in Fig.~\ref{fig:tetraquarks}.
In what follows, we will limit the  discussion of the implications of the  model to $S$-wave and $P$-wave tetraquark states only~\cite{Cleven:2015era}.
Since the isospin singlet and triplet are almost degenerate, there should be in total 24 $S$-wave tetraquark states even without
considering radial excitations, which are expected to be about 400 MeV heavier than the corresponding ground states~\cite{Maiani:2014aja}.
Among the ground states, the authors of  Ref.~\cite{Maiani:2014aja} identified the $X(3872)$ as the isoscalar
  $1^{++}$ ground state and $Z_c(3900)$ and $Z_c(4020)$ as the iso-triplet
  $1^{+-}$ ground states, respectively. One may also assign $X(3915)$ and $X(3940)$
  as $0^{++}$ and $2^{++}$,
  respectively~\cite{Maiani:2014aja}, although there are sizeable deviations between the predicted and the empirical values
  of the masses, cf. Fig.~\ref{fig:tetraquarks}.
  Therefore, at least 15 more $S$-wave tetraquarks still needs to be observed or
  identified.

For the $P$-wave tetraquarks, there are four iso-singlet $1^{--}$ states
without radial excitations.
Three of them are identified as the $Y(4008)$, $Y(4260)$ and $Y(4630)$, and
  the other one was identified as one of two structures, called $Y(4220)$ and
  $Y(4290)$ observed in $e^+e^-\to h_c\pi^+\pi^-$ ~\cite{Maiani:2014aja}.
  The states $Y(4360)$ and $Y(4660)$ were assigned to be the radial excitations of
the $Y(4008)$ and $Y(4260)$\cite{Maiani:2014aja}. Thus, besides the first
  radial excitations, only 6 of 112 (28 if considering only the iso-singlet
  ones) $P$-wave tetraquark states have candidates so far.
 As shown in Fig.~\ref{fig:tetraquarks} (b), there are two exotic quantum numbers, i.e. $0^{--}$ and $1^{-+}$,
  and two $0^{-+}$ states which might mix with each other. Since the mass decreases with the increasing $J$,
  a rather light charmonium(-like) state with $J=3$ is the particular feature of the
  discussed tetraquark picture\cite{Maiani:2014aja}.

\subsection{Hadronic molecules}
\begin{figure}[tbh] \centering
\includegraphics[width=0.9\linewidth]{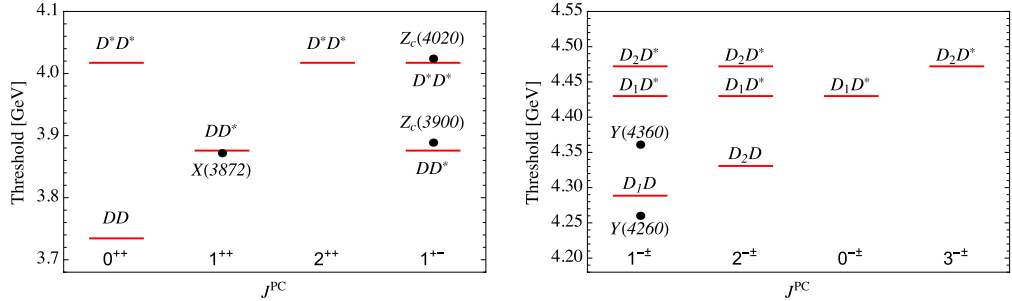}
\caption{The two-body thresholds in the charmonium mass range potentially related to the formation of hadronic molecules with fixed quantum numbers. The left and right panel are for $\frac 12+\frac 12$ and $\frac 12+\frac 32$ hadronic molecules, respectively.}
\label{fig:molecules}
\end{figure}

A hadronic molecule is an extended object which is composed of two or more
narrow hadrons via their nonperturbative interactions.
Because of their relatively narrow widths the ground state  hadrons have a long
enough life time to form a bound state.
Since some of the $XYZ$ states are close to $S$-wave thresholds and strongly
couple to the corresponding continuum states, an interpretation as hadronic
molecules appears quite natural.
In this contribution, we focus on the interaction of the members of two narrow
charmed meson doublets with one of them contain a $\bar c$ quark.
Those are characterized  by the quantum numbers of their light degrees of
freedom, namely
  $s_\ell^P=\frac12^-$ for the  doublet  $(D, D^*)$  and $s_\ell^P=\frac32^+$
  for the doublet $(D_1(2420), D_2(2460))$.
In particular, we discuss the $\frac12 + \frac 12$ and $\frac12 + \frac 32$
hadronic molecules.
There are two kinds of interaction between them: the long-ranged one-pion
exchange and short-ranged interactions.
The latter  need to be fixed by  a fit to experimental data.
Since there are not enough data available, for now we mostly restrict the
discussion to qualitative statements based on  the pion exchange potential.
As a result, instead of a prediction for the spectrum we show the potentially
relevant thresholds for the $\frac12+\frac 12$ molecules and $\frac 12+\frac 32$
molecules (cf. Fig.\ref{fig:molecules}).
However, this already allows us to highlight some striking features of hadronic
molecules which can be used to distinguish them from the hadro-charmonium and
tetraquark scenarios.

There is one exception to this lack of predictive power: as discussed in \cite{Guo:2013sya}, to leading order the potentials for $1^{++}$ and $2^{++}$ are identical, which means an iso-singlet hadronic molecule with quantum number $2^{++}$ near the $D^*D^*$ threshold would have a large probability to exist if the $X(3872)$ is assigned as an iso-singlet $DD^*$ molecule.

As is well-known the one-pion exchange potential has different signs in the
iso-singlet and iso-triplet channels.
Hence it is natural that an iso-triplet state does not exist if the $D$ meson
pair forms an iso-singlet and vice versa.
This means we do not expect any iso-triplet state with quantum numbers
$1^{++}$ or $2^{++}$. Due to the same reason, once $Z_c(3900)$ and $Z_c(4020)$
are assigned as isovector $D D^*$ and $D^*D^*$ hadronic molecules, one would not
expect to find isoscalar hadronic molecules with quantum number $1^{+-}$ at the
similar mass region.
For the $0^{++}$ channel the absence of the one-pion exchange in the $D D$
diagonal potential prevents us from speculating about the existence of molecular
structures with these quantum numbers.

For the $\frac12+\frac32$ system,  because there are two kinds of one-pion
exchange contributions, i.e. $t$-channel and $u$-channel exchange diagrams.
They depend on different products of unknown coupling constants, and thus no
strong conclusion can be drawn regarding the existence or non-existence of these
states.
Yet, one may still have reasonable speculations without dynamical analysis.
For instance, since hadronic molecules exist near the $S$-wave thresholds, the
lightest $\frac12+\frac32$ molecular state is expected be close to the $D_1 D$
threshold (cf. Fig.~\ref{fig:molecules}), and have quantum numbers of either
$1^{--}$ or $1^{-+}$. If a $0^{-\pm}$ molecule exists, it should be around the
$ D_1D^*$  threshold which is similar to that in the tetraquark picture and very
different from that in the hadro-charmonium picture where the $0^{-+}$ state is
expected to be the lightest one.
Another way to distinct between molecular and tetraquark scenarios is the
location of the  $J=3$ state. It will be the lightest state in the tetraquark
picture, but close to the $D_2D^*$ threshold (cf. Fig.~\ref{fig:molecules}) in
the molecular picture.

\section{Summary and Outlook}

In this contribution, we demonstrate that a narrow pronounced peak in the
elastic channel cannot be produced by purely kinematic effects.
A consistent treatment of these narrow structures necessarily calls for poles in
the $S$-matrix which correspond to physical states.
To further explore the nature of these $XYZ$ states, namely whether they are
hadro-charmonia, tetraquarks or hadronic molecules, we study their spectra in
those different scenarios using mainly HQSS and identify their distinctive
features.
The prominent components of
those puzzling states need to be identified by further experiments and detailed
analysis, which will deepen our understanding of QCD in the nonperturbative
regime.

\section{Acknowledgements}

We acknowledge Ulf-G. Mei{\ss}ner for his useful comments. This work is
supported in part by DFG and NSFC through funds provided to the Sino-German CRC
110 ``Symmetries and the Emergence of Structure in QCD'' (NSFC Grant No.
11261130311) and by NSFC (Grant No.~11165005).

\end{document}